# Displacement field calculation of large-scale structures using computer vision with physical constraints

Yapeng Guo[1], Peng Zhong[1], Yi Zhuo[2], Fanzeng Meng[2], Hao Di[2], Shunlong Li[1*]

[1] *School of Transportation Science and Engineering, Harbin Institute of Technology, Harbin, 150009, China*

2 *China Railway Design Corporation, Tianjin, 300142, China*

**Abstract**: Because of the advantages of easy deployment, low cost and non-contact, computer vision-based structural displacement acquisition technique has received wide attention and research in recent years. However, the displacement field acquisition of large-scale structures is a challenging topic due to the contradiction of camera field of view and resolution. This paper presents a large-scale structural displacement field calculation framework with integrated computer vision and physical constraints using only one camera. Firstly, the full-field image of the large-scale structure is obtained by processing the multi-view image using image stitching technique; secondly, the full-field image is meshed and the node displacements are calculated using an improved template matching method; and finally, the non-node displacements are described using shape functions considering physical constraints. The developed framework was validated using a scaled bridge model and evaluated by the proposed evaluation index for displacement field calculation accuracy. This paper can provide an effective way to obtain displacement fields of large-scale structures efficiently and cost-effectively.

---

*Corresponding author.
 Email-address: lishunlong@hit.edu.cn (S. Li)

# 1. Introduction

Displacement information, especially structural displacement field information, is essential for accurate service condition assessment of large-scale engineering structures. The analysis of structural displacements allows the sensing of local or global damage to the structure, and is an important basic response type for structural modal identification, damage identification and safety assessment[1-4]. Therefore, many types of displacement sensing devices have been used in structural health monitoring systems or structural inspection process[5-10], which can be divided into contact-based and non-contact-based.

The most commonly used contact-based sensors mainly include linear variable differential transformer (LVDT) and GPS [11,12]. The limitation of these two methods is the need of complex installations close to the structure. LVDT measures the relative displacement of a structure to a stationary point, meaning that a stationary point close to the structure must be set free from any vibration, which is often difficult to find in practice. GPS calculates the displacement by measuring the coordinates of the equipment installed on the structure. In addition to being expensive, the measurement accuracy is very limited, usually ± 1.5cm in the horizontal direction and ± 2cm in the vertical direction. The non-contact-based displacement sensor has the advantage of remote measurement without the need for complex installation on the structure. As a widely used non-contact sensor, laser displacement sensor needs a stationary point like LVDT, but the measurement distance cannot be too far due to the limitation of laser intensity.

As another major non-contact device, vision-based displacement sensors have received extensive attention from researchers due to their low cost, long measurement distance, multiple measurement points, and high measurement accuracy [13,14]. Vision-based structural displacement measurement calculates the displacement of a structure by comparing the position changes of the same pixels in different frames of the time-series images (video) of the structure remotely captured by cameras. Template matching and its variants are often used to find the location of the same pixel in different frames, namely tracking. To reduce the difficulty of tracking, initial research has been done to increase the recognizability of the appearance by matching artificial markers or targets installed at the location of the structure. With the improvement of the complexity and robustness of algorithms, satisfactory accuracy can be obtained by directly using the natural texture of the structure for tracking. Feng and Feng [15] proposed a multi-point simultaneous extraction method of structural displacement based on two improved template matching methods (using only one camera). Aoyama et al. [16] developed a multiple vibration distribution synthesis method to perform modal analysis on large-scale structures by using a multithread active vision system and a galvanometer mirror to perform quasi-real-time observation of multiple points of the structure. Luo et al. [17] proposed a set of image processing algorithms after analyzing the problems in practical outdoor applications, including the use of gradient-based template matching method, sub-pixel method and camera vibration elimination method. Due to the limited field of view of a single camera, Lydon et al. [18] developed multi-point displacement measurement system for large-scale structures using multiple time-synchronized wireless cameras and successfully applied it to actual bridge displacement measurements. Xu et al. [19] presented a multi-point displacement extraction method for real cable-stayed pedestrian bridges using consumer-grade cameras and computer vision algorithms. To solve the problem that the traditional image methods are not robust enough to the change of ambient light intensity, Song et al. [20] proposed to use fully convolutional network and conditional random field to segment the structural part from the image to extract the multi-point displacement combined with the digital image correlation method. These methods are aimed at extracting the displacement of one or several positions of the structure to be measured, that is, local displacement measurement.

Compared with local displacement, full-field displacement information of structures can provide more abundant structural state information for finite element model updating, material performance parameter identification, and structural condition assessment [21]. In addition, the visual sensor has the advantage of large-scale dense sensing, so it is more meaningful to conduct vision-based structural full-field displacement measurement. Compared with previous tracking methods such as digital image correlation or template matching, the phase-based method fits the full-field information acquisition and can obtain subpixel displacement measurement accuracy. Shang and Shen [22] proposed to use the phase-based optical flow method to obtain the full-field vibration map of the structure and use the motion magnification technology to identify the modal parameters. Yang et al. [23,24] used the physics-guided unsupervised machine learning vision method to identify the full-field vibration modes of stayed cables, and for the vibration structure with large rigid body displacement, a vision-based simultaneous identification method of rigid body displacement and structural vibration was also proposed, which was verified on the laboratory model. Narazaki et al. [25,26] developed a vision-based algorithm for measuring the dense three-dimensional displacement field of structures, and optimized the algorithm parameters using a laboratory truss model. Bhowmick and Nagarajaiah [27-29] proposed to use the continuous edges of the structure in the image as texture features and combine the optical flow method to extract the full-field displacement of the structure, and verified it on the three-layer steel frame model in the laboratory. To further simplify the measurement process of structural full-field displacement, Luan et al. [30] developed a deep learning extraction framework for structural full-field displacement based on convolutional neural networks, which realized real-time measurement of full-field subpixel displacement and verified it on a laboratory model. These studies have greatly promoted the development of structural full-field displacement measurement. However, since most of the work has been verified by small-scale laboratory models, a main problem in actual large-scale structural full-field displacement measurement is not involved: full-field structure image acquisition.

In vision-based structural displacement measurement, each pixel can be regarded as a sensor, and the actual distance it represents is the resolution of the measurement system. Although the accuracy can be further improved by means of subpixel technology, it can only be amplified by an extremely limited multiple. Therefore, obtaining a full-field image of the structure with sufficient resolution is the basis of full-field displacement measurement. Due to the small size of the laboratory model, the field of view of a camera can cover the entire structure with good resolution. But actual civil structures tend to be huge, if only one camera to shoot all the structure will result in extremely low resolution, and to maintain the resolution will result in the camera 's field of view is too small to cover the whole structure.

To solve the image acquisition and processing problem in full-field displacement measurement of large-scale structures, this paper presents a novel calculation framework of large-scale structural displacement field. To alleviate the contradiction between camera field of view and resolution, image stitching technology based on multi-view images is proposed to generate large-scale structure full-field images. To improve the efficiency of displacement extraction and consider the physical rules, the node and non-node displacement extraction technology based on meshing and structural shape function is developed. The remainder of this paper is organized as follows. Section 2 describes the details of the proposed structural displacement field calculation framework. Section 3 illustrates the verification results and discusses the key parameters of the presented method. Finally, Section 4 concludes the study.

## 2. Structural displacement field calculation framework

The proposed calculation framework of structural displacement field is shown in Figure 1. Firstly, a camera set on the automatic rotation device is used to shoot the large-scale structure to obtain a multi-view structure image. Secondly, the full-field structure image is generated by using image stitching technology. Finally, the full-field image is discretized and meshed, the displacement at the node is calculated by the improved template matching method, and the displacement at the non-node is calculated by the shape function considering the physical rules, to obtain the displacement field of the large-scale structure.

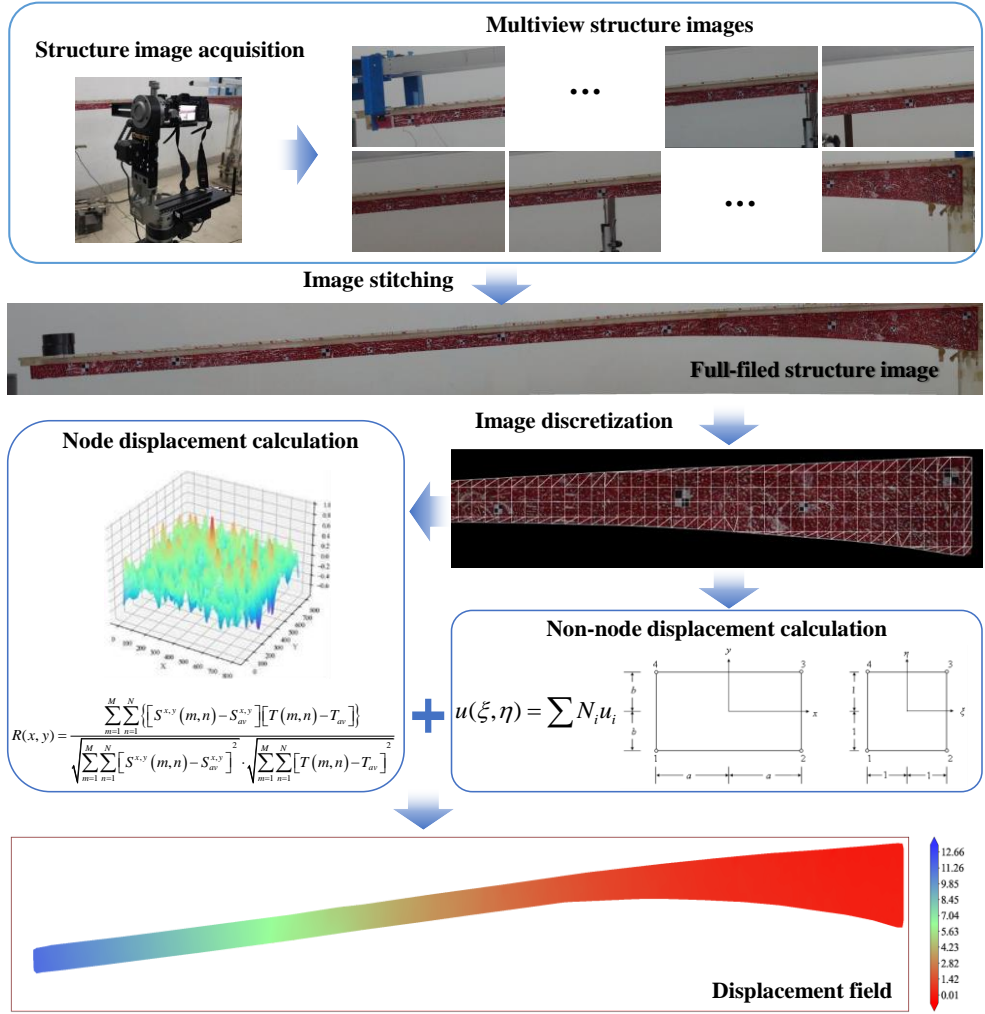

Figure 1. Overall framework of the proposed structural displacement field calculation method

## 2.1 Large-scale structure full-field image generation using image stitching

To obtain high-resolution full-field images of large-scale structures, this paper proposes a full-field image generation method that rotates and moves a single camera to capture multiple partial structure images and stitch them together. The proposed generation method is divided into three steps: (1) image preprocessing (used to solve the problem of inconsistent depth of field), (2) image registration (used to align and stitch multi-view images), (3) structure foreground segmentation (used to extract the structure in the full-filed image).

2.1.1 Image preprocessing

Multi-view imaging of large-scale structures by rotating the camera is usually convenient. However, it is accompanied by the fact that the same structure plane has different depth of field in different images (foreshortening effects) due to the angle problem during each imaging. Only unifying the structural planes in all images can ensure no distortion in the subsequent stitching process. This can be achieved by rotating the camera imaging plane around the intersection of the optical axis and the imaging plane to be parallel to the structure plane.

This paper proposes to use perspective transformation to re-project the original camera imaging surface to a new structure plane. The homography matrix is usually used to describe the transformation between such two-dimensional planes. The homography matrix can be solved by finding the coordinates of four points in the old and new images. Because the rotation vector and translation vector can be measured when the camera takes multi-view images, the corresponding new coordinates of the four points can be calculated based on these two vectors to achieve image preprocessing.

2.1.2 Image registration

The preprocessed image needs to be stitched into a full-field image after removing the interference factors. Usually, the feature points of each image are calculated based on feature point detection algorithms, and the same feature points in the two images are matched with each other to calculate the homography matrix representing the transformation. Due to the different external conditions during image shooting, the overlapping areas of adjacent images will also be different, the image fusion method is needed to make the stitching effect more natural.

In this paper, the scale-invariant feature transform (SIFT) algorithm is used to detect local feature points in the image. SIFT features still show good feature detection results and strong robustness even in complex environments such as scale changes, image rotation and brightness changes. The SIFT algorithm will simultaneously generate the coordinates of the feature points and the corresponding descriptors. For two feature points with descriptors of $\boldsymbol{R} = (r_1, r_2, \ldots, r_n)$ and $\boldsymbol{S} = (s_1, s_2, \ldots, s_n)$, the Euclidean distance is calculated to evaluate their similarity.

The Fast Library for Approximate Nearest Neighbors (FLANN) is used to match the feature point sets in two adjacent images. Using the K-D (k-dimensional) tree, all the feature points in the image are divided into left and right sub-tree spaces according to the root nodes of different dimensions. Then the root nodes are determined in the sub-tree space, and the space is divided again until the space is empty, that is, all the feature points are divided. After using FLANN, there will inevitably be mismatches. If it is included in the calculation of homography matrix, there will be obvious errors in splicing. In this paper, Random Sample Consensus (RANSAC) algorithm is used to filter and only retain the correct matching feature points. The direct average fusion method is used to recalculate and replace the pixel value of the overlapping area using the average pixel value of adjacent images.

2.1.3 Structure foreground segmentation

Structure full-field image includes not only the structure itself, but also inevitably includes sensors, bearings, background interference and so on. However, the displacement field of the structure is only generated in the structure itself, and the other objects must be removed. Therefore, this paper uses the GrabCut algorithm to extract the foreground that contains only structures.

The GrabCut algorithm is an improvement of the GraphCut algorithm, which is mainly reflected in the following aspects. First, simplify the user interaction operation, only need to roughly mark the rectangular box containing the foreground object, and the outside of the box is the background. Second, instead of gray histogram, Gaussian mixture model (GMM) is used to estimate the probability of pixels belonging to foreground and background, and the calculation results are more reliable and accurate. Third, segmentation is not a one-time completion, through continuous iteration, update calculation parameters, so that the image segmentation quality is improved. After users mark the bounding box, all pixels outside the box belong entirely to the background, and the pixels inside the box may belong to both the foreground and the background. Therefore, it is only necessary to segment the connection relationship between pixels in the box to separate the foreground and background.

## 2.2 Structure image discretization and displacement field calculation

Based on the finite element concept in the physical model, the proposed computer vision-based structural displacement field calculation with physical constraints can be divided into three steps: firstly, the continuous structural foreground image is discretized, and the mesh is drawn within the foreground image. The structure is divided into several regions by using the mesh. Secondly, the displacement of structural grid nodes is calculated based on the improved template matching method. Thirdly, the shape function is constructed to establish the relationship between the displacement at the grid nodes and the displacement at the non-nodes in the grid, the displacement of the nodes is transferred to the non-nodes by the shape function to generate a complete displacement field. Because the shape function is a continuous function, the generated displacement field is also continuous.

2.2.1 Structure image discretization

After determining the mesh size, the horizontal and vertical lines of the mesh are drawn on the full-field image. The grid has not abandoned the background part, which is a full-size grid, as shown in Figure 2. It is a regular rectangular grid discretization of the whole image.

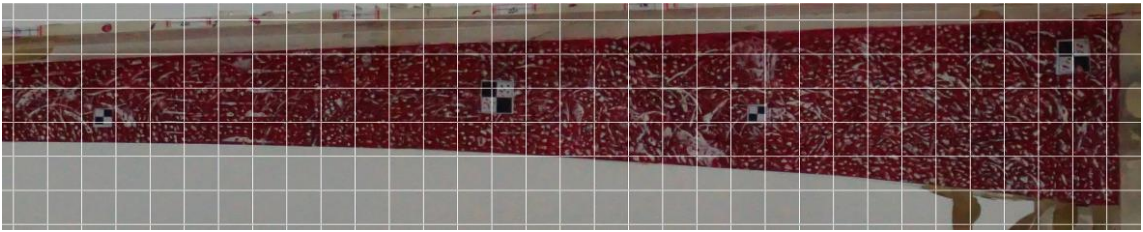

Figure 2. Full-size grid of the full-field structure image

Like foreground extraction, it is also necessary to generate the boundary of the displacement field on the full-size grid. First, the binary threshold processing is performed on the structure foreground extraction image. The background is set to 0 and the structure is set to 1 to form a binary foreground image, called a mask. To retain as many pixels as possible and ensure that the boundary of the foreground extraction has some surplus pixel space, the structure element with a kernel of $3 \times 3$ is used to expand the morphological processing of the binary image, and finally the expanded structure image mask is formed. Each pixel of the mask and the corresponding pixel of the full-size grid map are bitwise and calculated, that is, 1 & 1 = 1, 0 & 1 = 0, to retain the grid within the structural range. The endpoints of the above dividing line are connected to form a grid with boundary. The grid division has been initially formed, as shown in Figure 3.

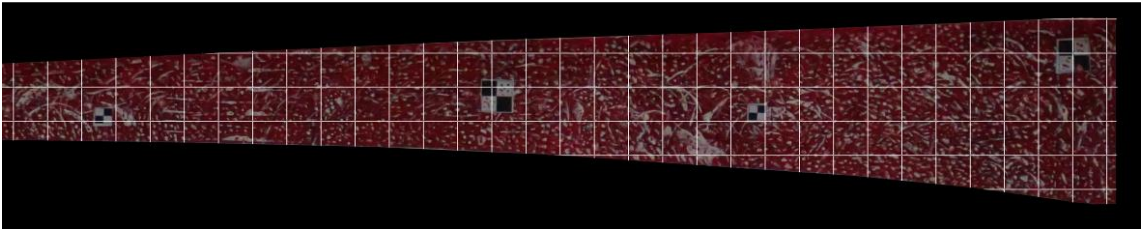

Figure 3. Structure region grid of the full-field structure image

The grid that does not contact the boundary is a regular rectangular grid, but the grid rules that contact the boundary are different, and further fine division needs to be completed. The grid shapes at the edge are trapezoid, pentagon and triangle. To unify shape and refinement, the trapezoid and pentagon are divided into several triangles. So far, the division of the entire displacement field grid is completed, and the division unit has only rectangles and triangles (as shown in Figure 4), which is convenient for subsequent construction of shape functions according to the unit shape.

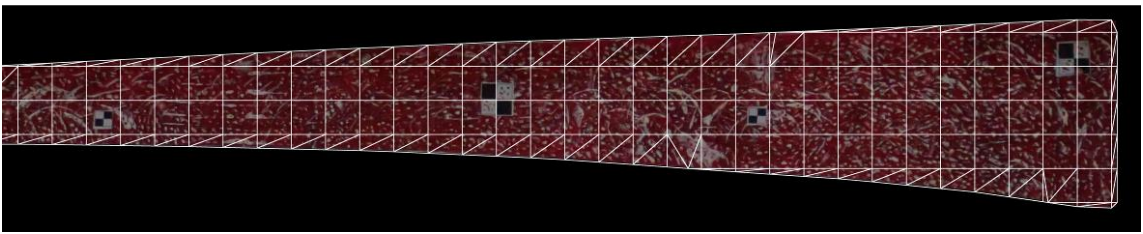

Figure 4. Final grid of the full-field structure image

2.2.2 Node displacement calculation using template matching

Template matching belongs to digital correlation technology. The principle is shown in the figure. First determine the template image, generate a window of the same size as the template image, and traverse the window from the upper left corner to the lower right corner in the image to be matched. Through correlation calculation, the correlation coefficient between all window images and template images is obtained, and a matrix with integer pixels is generated. The window corresponding to the peak of the correlation coefficient is the location of the template image in the image to be matched.

The core of template matching lies in the correlation calculation. The normalized sum of squared differences (NSSD) is based on the sum of squared differences to calculate the difference between the gray values of the template image and the window image pixel points. The method is simple. Although the gray value can be normalized, the effect of white noise can be weakened, but the amount of computation is too large and it is very sensitive to illumination. The normalized cross-correlation function (NCC) is to multiply the pixel values of points on the same pixel coordinates of two images of equal size, and then compare them with the square sum root of the pixel values of all pixels of the two images. Because in the calculation process, all the pixel values are squared and then the root number is processed, which can well weaken the influence of image white noise. Although the normalized cross-correlation function can resist white noise, it cannot solve the problem of brightness inconsistency. When the same pattern in different brightness environment, the normalized cross-correlation function calculated similarity is very low. Therefore, to solve this shortcoming, the zero-mean normalized cross-correlation function (ZNCC) is improved based on the normalized cross-correlation function (shown in Equations (1) and (2)). In the calculation process, the pixel value of all pixels is subtracted from the average value of the image pixel value, thereby weakening the influence of the image brightness on the calculation result.

$$R(x,y) = \frac{\sum_{m=1}^{M}\sum_{n=1}^{N}\left\{\left[S^{x,y}(m,n) - S_{av}^{x,y}\right]\left[T(m,n) - T_{av}\right]\right\}}{\sqrt{\sum_{m=1}^{M}\sum_{n=1}^{N}\left[S^{x,y}(m,n) - S_{av}^{x,y}\right]^{2}} \cdot \sqrt{\sum_{m=1}^{M}\sum_{n=1}^{N}\left[T(m,n) - T_{av}\right]^{2}}} \quad (1)$$

$$S_{av}^{x,y} = \frac{1}{MN}\sum_{m=1}^{M}\sum_{n=1}^{N}S^{x,y}(m,n), T_{av} = \frac{1}{MN}\sum_{m=1}^{M}\sum_{n=1}^{N}T(m,n) \quad (2)$$

2.2.3 Non-node displacement calculation using shape function

For non-node displacement, it is necessary to transfer the node displacement to each grid element by means of the idea of finite element shape function in the physical model. Triangular and rectangular element shape functions are constructed according to the element types of the previous mesh generation results.

(1) Rectangular bilinear element shape function

The rectangular element has four nodes, so the analysis model of rectangular bilinear element is adopted, and there are 8 node displacement parameters. To simplify the results, the rectangular coordinates are transformed into regular coordinates for analysis by means of coordinate transformation, as shown in Figure 5.

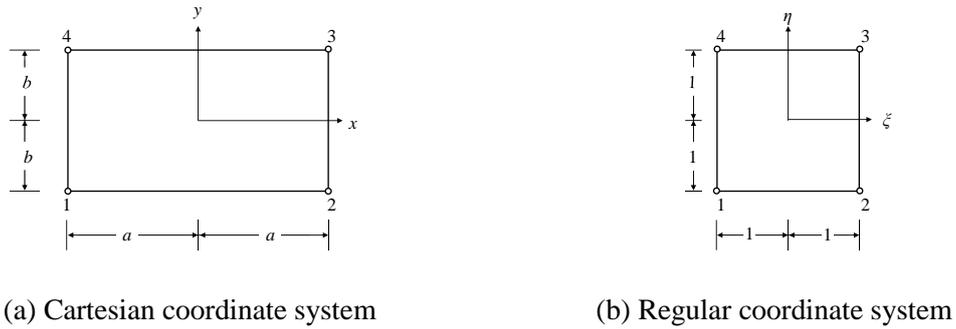

(a) Cartesian coordinate system  (b) Regular coordinate system

Figure 5. Rectangular bilinear element

In the regular coordinate system, the boundary line equation of rectangular element is:

$$\begin{cases} \eta + 1 = 0 \\ \xi - 1 = 0 \\ \eta - 1 = 0 \\ \xi + 1 = 0 \end{cases} \tag{3}$$

According to the property that the shape function $N_i$ is 1 at node $i$ and 0 at other points, the following equation can be acquired:

$$\begin{cases} N_i(\xi_i, \eta_i) = 1 \\ N_i(\xi_j, \eta_j) = 0, i \neq j \end{cases} \tag{4}$$

The shape function of each point can be set as:

$$\begin{cases} N_1 = \alpha(\xi - 1)(\eta - 1) \\ N_2 = \beta(\xi + 1)(\eta - 1) \\ N_3 = \gamma(\xi + 1)(\eta + 1) \\ N_4 = \delta(\xi - 1)(\eta + 1) \end{cases} \tag{5}$$

Substituting Equation (4) into Equation (5) yields:

$$\alpha = -\beta = \gamma = -\delta = \frac{1}{4} \tag{6}$$

Given the displacement $u_i$ of each node $i$, the displacement function of each point in the element is:

$$u(\xi, \eta) = \sum N_i u_i \tag{7}$$

(2) Triangular element shape function

Since the shape function of triangular element in rectangular coordinate system will be more complex, the area coordinate is introduced for analysis.

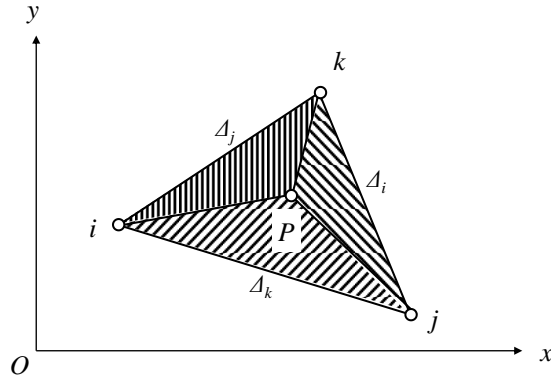

Figure 6. Area coordinate of triangular element

As shown in Figure 6, $i$, $j$, $k$ are triangular element nodes, and $P$ is any point in the element, which is connected to each node. Then $\Delta_{ijk}$ is divided into three parts, which are denoted as:

$$\begin{cases} \Delta_i = \Delta Pjk \\ \Delta_j = \Delta Pki \\ \Delta_k = \Delta Pij \\ \Delta = \Delta_i + \Delta_j + \Delta_k = \Delta ijk \end{cases} \tag{8}$$

The position of point $P$ can be represented by rectangular coordinates, and can also be represented by $\Delta_i$, $\Delta_j$, $\Delta_k$.

$$L_l = \frac{\Delta_l}{\Delta} \quad (l = i, j, k) \tag{9}$$

Then the position of point P can also be determined by $L_l$, which is called area coordinate.

Let the point $P$ coordinate be $(x, y)$, then the area divided into three parts is:

$$\Delta_i = \frac{1}{2}\begin{vmatrix} x & y & 1 \\ x_j & y_j & 1 \\ x_k & y_k & 1 \end{vmatrix} \quad i \to j \to k \to i \tag{10}$$

The unit area is:

$$\Delta = \frac{1}{2}\begin{vmatrix} x_i & y_i & 1 \\ x_j & y_j & 1 \\ x_k & y_k & 1 \end{vmatrix} \tag{11}$$

Therefore, the area coordinate is:

$$L_i = \frac{\Delta_i}{\Delta} = \frac{1}{2\Delta}\left( x\begin{vmatrix} y_j & 1 \\ y_k & 1 \end{vmatrix} - y\begin{vmatrix} x_j & 1 \\ x_k & 1 \end{vmatrix} + 1\begin{vmatrix} x_j & x_j \\ x_k & x_k \end{vmatrix} \right)$$
$$= \frac{1}{2\Delta}(a_i + b_i x + c_i y) \quad (i \to j \to k \to i) \tag{12}$$

$$\begin{cases} a_i = x_j y_k - x_k y_j \\ b_i = y_j - y_k \\ c_i = x_k - x_j \end{cases} \quad (i \to j \to k \to i) \tag{13}$$

According to the properties of shape functions, $L_i$, $L_j$ and $L_k$ are the shape functions of triangular elements. The displacement of each point inside the triangular element is calculated by Equation (7).

## 3. Experimental verification results and discussions

### 3.1 Full-field image generation results

The bridge model used in this test is a side span of organic glass three-span continuous beam. To reduce the structural stiffness, the side span pier was removed and turned into a cantilever beam structure (2644mm long). The camera is consumer-grade (Sony α-6000) with a resolution of 6000 × 4000. The structure was photographed from left to right using a fully automatic rotating pan-tilt control platform, as shown in Figure 7. The rotation speed was 2 degrees per second, and a total of 20 structural images were obtained. To verify the accuracy of the proposed method for displacement measurement, artificial targets of known sizes (20mm × 20mm) were placed at 200mm, 400mm, 800mm, 1300mm, 1800mm and 2200mm from the beam end, two LVDT displacement sensors were installed 400 mm and 1300 mm, respectively.

Four loading cases were tested (different loadings placed at the cantilever end to excite different displacement fields of the structure, shown in Figure 8).

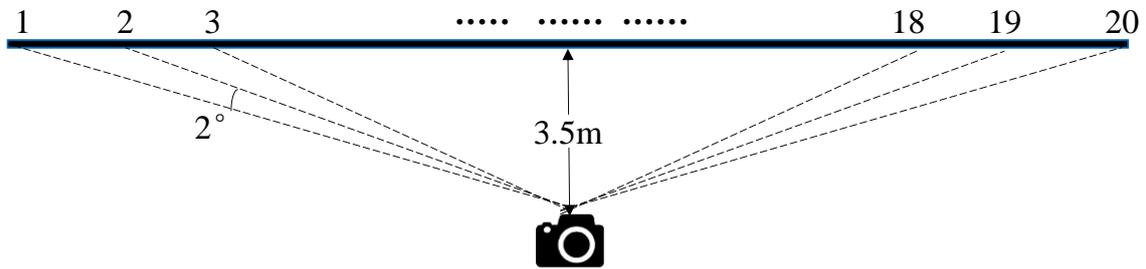

Figure 7. Schematic diagram of the rotation shooting process

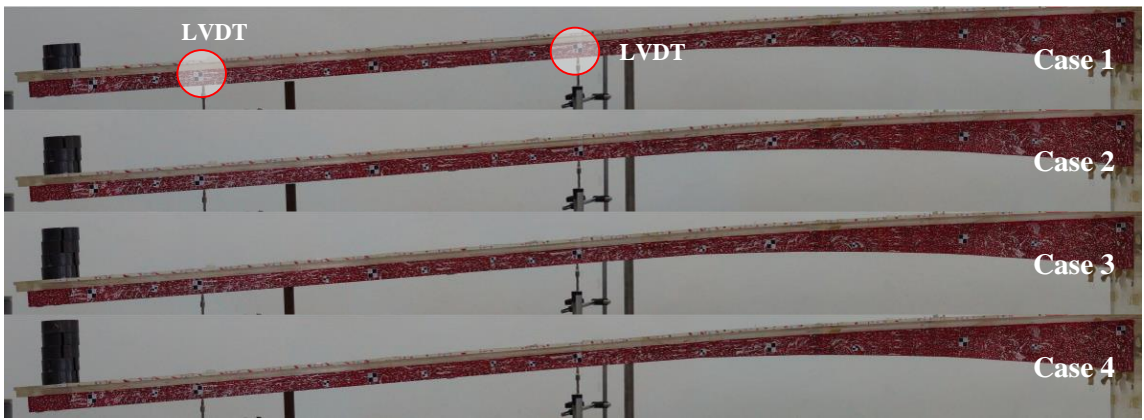

Figure 8. Four different loading cases

The pre-processing method was used to process the obtained 20 images to eliminate the foreshortening effect. The full-field image of the structure obtained by image stitching is shown in Figure 9. The conversion coefficient represents the actual distance represented by a pixel in the image. According to the artificial target, the conversion coefficient error of each part of the whole field image is calculated to be within 1%, which shows the effectiveness of the preprocessing algorithm. The structure foreground segmentation result is shown in Figure 10.

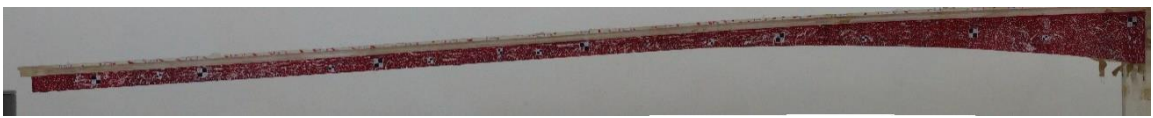

Figure 9. Full-field image of the structure obtained by image stitching

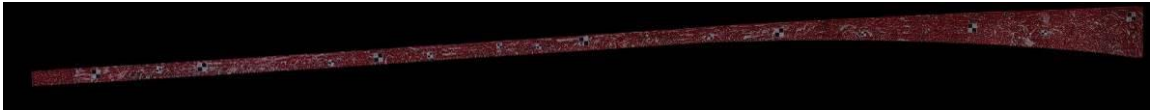

Figure 10. Structure foreground segmentation result

## 3.2 Displacement field calculation results

In the generated full-filed image, the lowest of the bridge structure is about 500 pixels, and the highest is about 1600 pixels. The size of the template in template matching technique is set to 81 × 81, and the mesh size is set to 400 × 400. With the initial state image as a reference, the proposed method was used to calculate the structural displacement field for four loading cases, and the results are shown in Figure 11-Figure 14. The calculated structural displacement field conforms to the law of structural mechanics and the displacement change is continuous, which can preliminarily validate the feasibility of the proposed method.

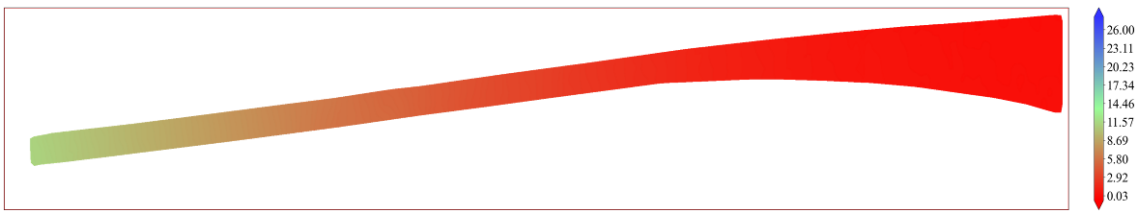

Figure 11. Calculated structural displacement field for Case 1

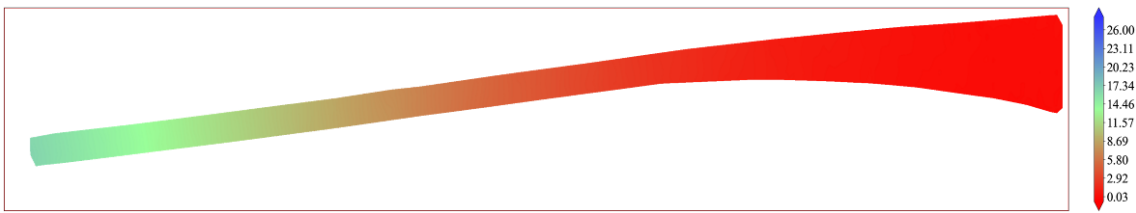

Figure 12. Calculated structural displacement field for Case 2

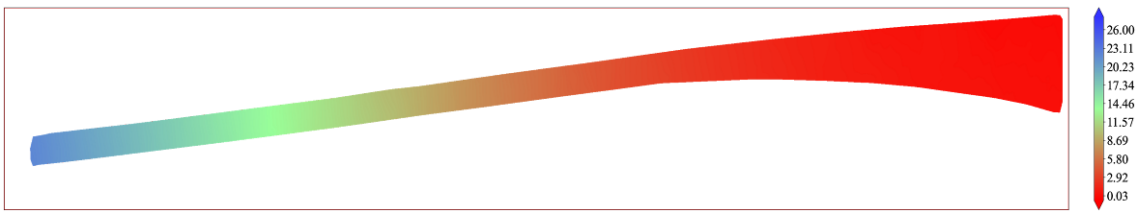

Figure 13. Calculated structural displacement field for Case 3

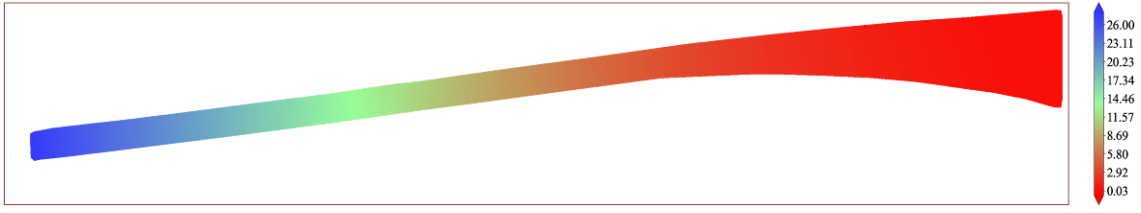

Figure 14. Calculated structural displacement field for Case 4

To quantitatively evaluate the accuracy of the structural displacement field calculated by the proposed method, the corresponding finite element model was established and updated according to the geometric size and material properties of the bridge model used in the experiment (the objective is the displacement difference at the position of LVDT), as shown in Figure 15. Considering the displacement field dimension calculated by the proposed method, the corresponding displacement fields of different load cases generated by the finite element model are extracted.

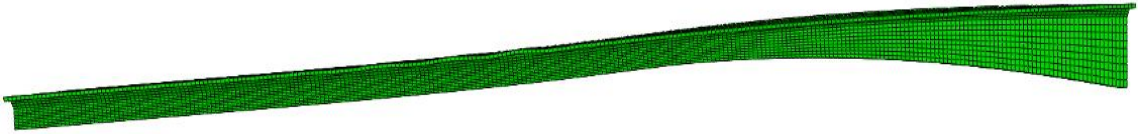

Figure 15. Finite element model of the employed bridge model

To evaluate the similarity of two displacement fields ($\mathbf{F}_1$, $\mathbf{F}_2$), the normalized correlation coefficient $\mathbf{R}$ ($\mathbf{F}_1$, $\mathbf{F}_2$) is proposed as the evaluation index. The calculation method is shown in Equation (14), where $M$ and $N$ are the number of rows and columns of the displacement field, and $\mathbf{F}$ ($m$, $n$) is the displacement value at the position of the displacement field ($m$, $n$). In addition, the deviation between data is also an important index to measure the difference of displacement field. The difference between two displacement field data is defined as the root mean square deviation $\mathbf{D}$ ($\mathbf{F}_1$, $\mathbf{F}_2$), and its calculation method is shown in Equation (15).

$$R(F_1, F_2) = \frac{\sum_{m=1}^{M}\sum_{n=1}^{N}\left[F_1(m,n) \cdot F_2(m,n)\right]}{\sqrt{\sum_{m=1}^{M}\sum_{n=1}^{N}\left[F_1(m,n)\right]^2} \cdot \sqrt{\sum_{m=1}^{M}\sum_{n=1}^{N}\left[F_2(m,n)\right]^2}} \quad (14)$$

$$D(F_1, F_2) = \sqrt{\frac{\sum_{m=1}^{M}\sum_{n=1}^{N}\left[(F_1(m,n) - F_2(m,n))^2\right]}{M \times N}} \tag{15}$$

Figure 16 shows the quantitative evaluation results of the structural displacement field under different load conditions calculated by the proposed method. The **R** values of the calculation results of the proposed method and the finite element model are higher than 0.9995, indicating that the overall trend of the calculated displacement field is similar to the real displacement field. The **D** values are all less than 0.304 mm, and the relative offset obtained by dividing the D value by the maximum displacement field of the four load cases is less than 1.2 %, which verifies the accuracy of the proposed method.

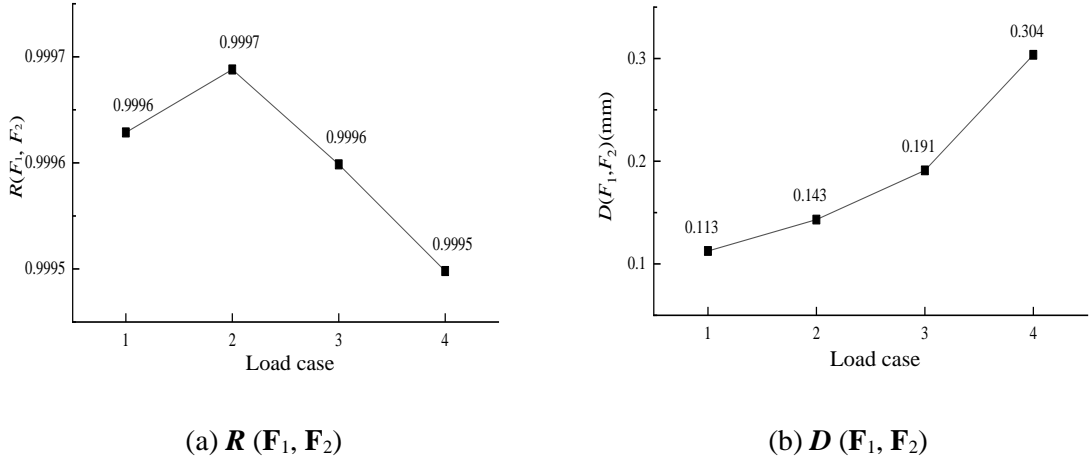

(a) $R(\mathbf{F}_1, \mathbf{F}_2)$  (b) $D(\mathbf{F}_1, \mathbf{F}_2)$

Figure 16. Quantitative evaluation results of the structural displacement field

### 3.3 The influence of different mesh size

As one of the important parameters of the proposed method, the mesh size affects the number of meshes and nodes in the full-field structural image. In the full-field bridge image of this experiment, the minimum structural height is only 500 pixels, so the mesh size should not be greater than 500. To study the influence of mesh size on the calculation of structural displacement field, five different meshes of 200 × 200, 250 × 250, 300 × 300, 350 × 350 and 400 × 400 were used to divide the image. The local meshes near the consolidation end is shown in Figure 17.

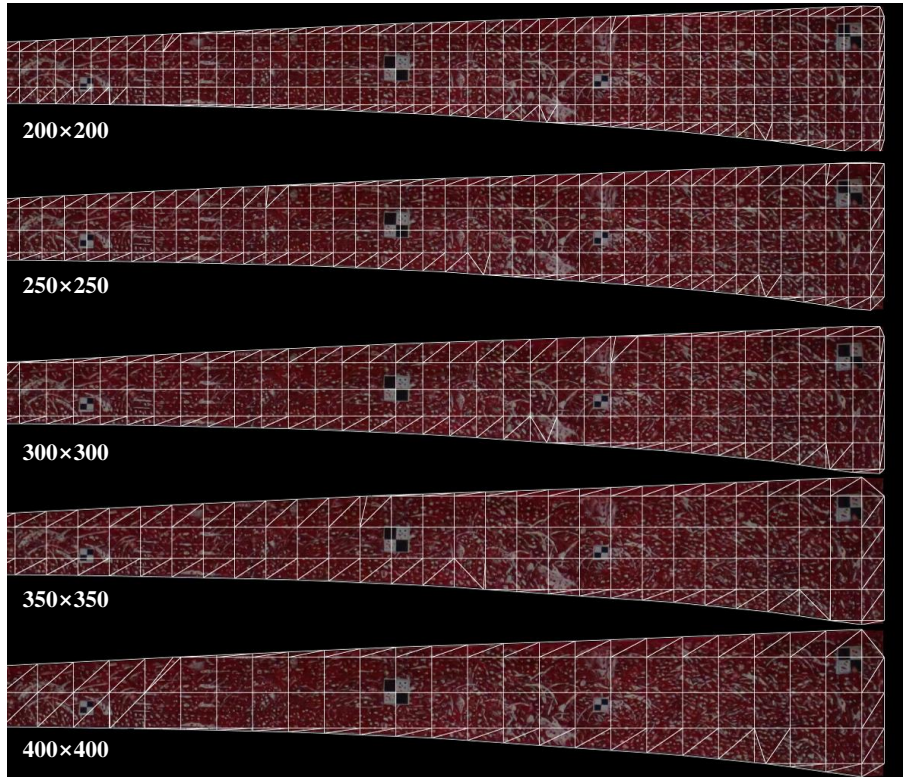

Figure 17. Local meshes near the consolidation end with different mesh sizes

The $R$ and $D$ values of the structural displacement field calculated with different mesh sizes are calculated respectively. The results are shown in Figure 18. The $R$ and $D$ values corresponding to the five mesh sizes are basically unchanged, which verifies that the mesh size has no significant effect on the calculation accuracy of the structural displacement field. However, considering that the larger the mesh size, the fewer the number of generated mesh and nodes, that is, the higher the calculation efficiency, it is recommended that the mesh size can be as large as possible within the allowable range.

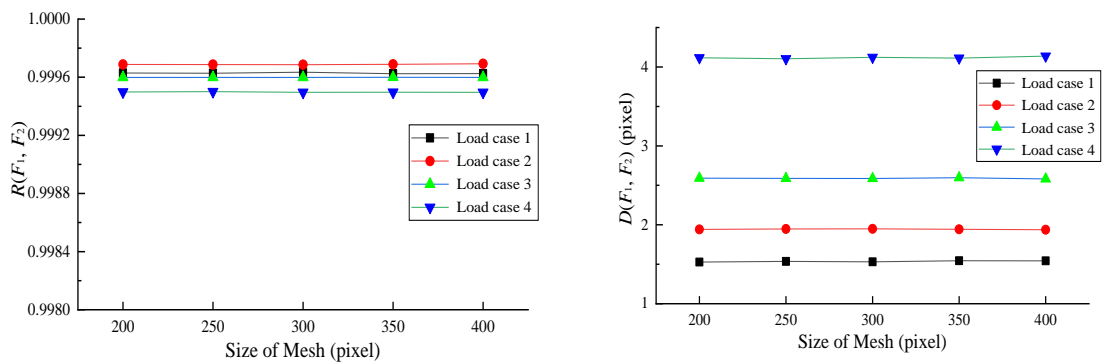

(a) $R$ ($F_1$, $F_2$)                 (b) $D$ ($F_1$, $F_2$)

Figure 18. Quantitative evaluation results with different mesh sizes

## 4. Conclusions

In this paper, a displacement field calculation framework of large-scale structures based on computer vision with physical constraints is proposed. Only a single camera is used to solve the contradiction between imaging field of view and resolution, and the accurate acquisition of large-scale structural displacement field is realized. The specific conclusions can be drawn as follows: (1) It is feasible to use the camera set on the automatic rotating device to obtain high-resolution images of large-scale structures and then use image stitching technology to generate panoramic images of structures; (2) The laboratory bridge model is used to verify the proposed framework, and an updated finite element model is established for quantitative evaluation. The evaluation index R is greater than 0.9995, and the D value is less than 0.304 mm, which validate the accuracy of the proposed method; (3) The parameter sensitivity analysis of mesh size, one of the important parameters, is conducted. The mesh size has no significant effect on the accuracy, but considering the computational efficiency, the mesh size can take the upper limit of the allowable range.

In this paper, the displacement field of a large structure is obtained at a small hardware cost, but the limitation is that it can only be used for static or quasi-static deformation of the structure, and can not realize real-time calculation of the dynamic displacement field. This is because the use of automatic rotating device for image shooting requires time that cannot be ignored. Future work will focus on improving the workflow of the rotating device or using multi-lens cameras for simultaneous shooting to minimize image acquisition time and further improve the efficiency of the algorithm to achieve real-time acquisition of dynamic displacement fields.

## Declaration of competing interest

The author(s) declared no potential conflicts of interest with respect to the research, authorship, and/or publication of this article.

# Acknowledgements

Financial support for this study was provided by the NSFC [Grant Nos. U22A20230 and 52278299], Fundamental Research Funds for Central Universities [Grant No. FRFCU5710051018] and China Railway Design Corporation R&D Program [2020YY240604].